\documentclass{midl} 

\usepackage{mwe} 
\usepackage{csquotes}
\usepackage{float}
\usepackage{adjustbox}
\usepackage{multirow}
\usepackage{xcolor}
\usepackage{caption}

\jmlrpages{}
\jmlryear{2025}

\title[How We Won the ISLES’24 Challenge by Preprocessing]{How We Won the ISLES’24 Challenge by Preprocessing}

\midlauthor{
\Name{Tianyi Ren}\midljointauthortext{Contributed equally}\nametag{$^{1}$}\orcid{0000-0001-9548-6645} \Email{tr1@uw.edu}\\
\Name{Juampablo E. Heras Rivera\midlotherjointauthor}\nametag{$^{1}$}\orcid{0000-0002-0205-6329} \Email{jehr@uw.edu}\\
\Name{Hitender Oswal\midlotherjointauthor}\nametag{$^{2}$}\orcid{0000-0002-4507-5466} \Email{hitender@uw.edu}\\
\Name{Yutong Pan}\nametag{$^{2}$} \Email{ypan4@uw.edu}\\
\Name{William Henry}\nametag{$^{1}$} \Email{whenry1@uw.edu}\\
\Name{Sophie Walters}\nametag{$^{3}$} \Email{smw59@uw.edu}\\
\Name{Mehmet Kurt}\nametag{$^{1}$}\orcid{0000-0002-5618-0296} \Email{mkurt@uw.edu}\\
\addr $^{1}$ Department of Mechanical Engineering, University of Washington, Seattle, WA, USA \\
\addr $^{2}$ Paul G. Allen School of Computer Science, University of Washington, Seattle, WA, USA \\
\addr $^{3}$ Department of Bioengineering, University of Washington, Seattle, WA, USA
}

\begin{document}

\maketitle

\begin{abstract}
Stroke is among the top three causes of death worldwide, and accurate identification of stroke lesion boundaries is critical for diagnosis and treatment. Supervised deep learning methods have emerged as the leading solution for stroke lesion segmentation but require large, diverse, and annotated datasets. The ISLES'24 challenge addresses this need by providing longitudinal stroke imaging data, including CT scans taken on arrival to the hospital and follow-up MRI taken 2-9 days from initial arrival, with annotations derived from follow-up MRI. Importantly, models submitted to the ISLES'24 challenge are evaluated using only CT inputs, requiring prediction of lesion progression that may not be visible in CT scans for segmentation. Our winning solution shows that a carefully designed preprocessing pipeline including deep-learning-based skull stripping and custom intensity windowing is beneficial for accurate segmentation.
Combined with a standard large residual nnU-Net architecture for segmentation, this approach achieves a mean test Dice of 28.5 with a standard deviation of 21.27.


\end{abstract}


\section{Introduction}
Stroke is among the top three causes of death worldwide, with ischemic strokes accounting for over 87\% of cases \cite{strokefax}. Rapid and accurate segmentation of acute ischemic stroke lesions from medical imaging is critical for accurate diagnosis and treatment planning. Supervised deep learning methods are the leading solution for stroke lesion segmentation but require large, diverse, and annotated datasets. The ISLES'24 \cite{isles24} challenge provides longitudinal imaging data of stroke patients, including acute non-contrast CT, CT angiography, CT perfusion scans, follow-up MRI acquired 2–9 days later, and ground truth hand-annotated lesion masks. The ground truth lesion annotations are obtained using follow-up MRI, typically showing lesion progression beyond what is visible on initial CT scans. Importantly, models submitted to the ISLES'24 challenge are evaluated using only CT inputs, requiring prediction of lesion progression that may not be visible in CT scans for segmentation.

The low signal-to-noise ratio of CT scans to ischemic insults leads to low senstivity to subtle lesions, making segmentation difficult. Unlike MRI-based tasks like BraTS \cite{brats}, where lesions are clearly visible, CT requires careful preprocessing to reveal informative features. The key insight of this work is that tailored preprocessing is necessary to increase lesion visibility and provide a clear learning signal for the segmentation model. Despite its simplicity, this significantly improves model performance and generalizability, and was key to our submission achieving first place in the ISLES'24 challenge.

\section{Dataset}
The ISLES'24 challenge dataset was used for model training and evaluation. This dataset contains imaging and tabular data from large vessel occlusion ischemic stroke patients acquired at two time points: admission (NCCT, CTA, CTP and perfusion maps), and follow-up 2 to 9 days later (DWI, ADC, and ground truth masks obtained from DWI). In total, the ISLES'24 dataset includes 250 sets of scans, with 150 sets for training (100 from the University Hospital of Munich and 50 from the University Hospital of Zurich), and 100 sets for testing (from undisclosed hospitals).

\section{Methods}
Our solution for the ISLES challenge includes a preprocessing step consisting of skull-stripping with SynthStrip \cite{synthstrip} and intensity windowing with custom windows, followed by a standard residual nnU-Net model \cite{nnunet} for segmentation. 

\subsection{nnU-Net}
Despite the emergence of models like Transformers \cite{hatamizadeh2021swin} and diffusion frameworks \cite{ren2024re}, CNNs based on the U-Net architecture \cite{unet} remain state-of-the-art for medical image segmentation \cite{nnunet_sota}. The nnU-Net framework \cite{nnunet} automates hyperparameter tuning by adapting a standard U-Net to the training data, often outperforming manually tuned and novel models \cite{rivera2024ensemble}. For our submission, we used a 3D large residual encoder nnU-Net, referred to as “nnU-Net ResEnc L” in the nnU-Net documentation. For training, a [56, 320, 256] patch size, Dice and cross-entropy loss, and the SGD optimizer (lr=0.01, momentum=0.99) were used.

\subsection{SynthStrip}
The scans in the ISLES'24 brain imaging dataset contain non-brain structures such as the skull and background artifacts, which can hinder model training. To address this, we apply SynthStrip, a deep-learning-based brain-extraction tool trained on diverse synthetic images. First, we applied SynthStrip on the non-contrast CT (NCCT) scans to obtain a brain masks. Then, we applied this brain mask to the other co-registered scans (CTP, CTA, etc.) to obtain skull-stripped versions of the data. 

\subsection{Intensity Windowing}

\begin{figure}[H]
    \centering
    \includegraphics[width=0.8\linewidth]{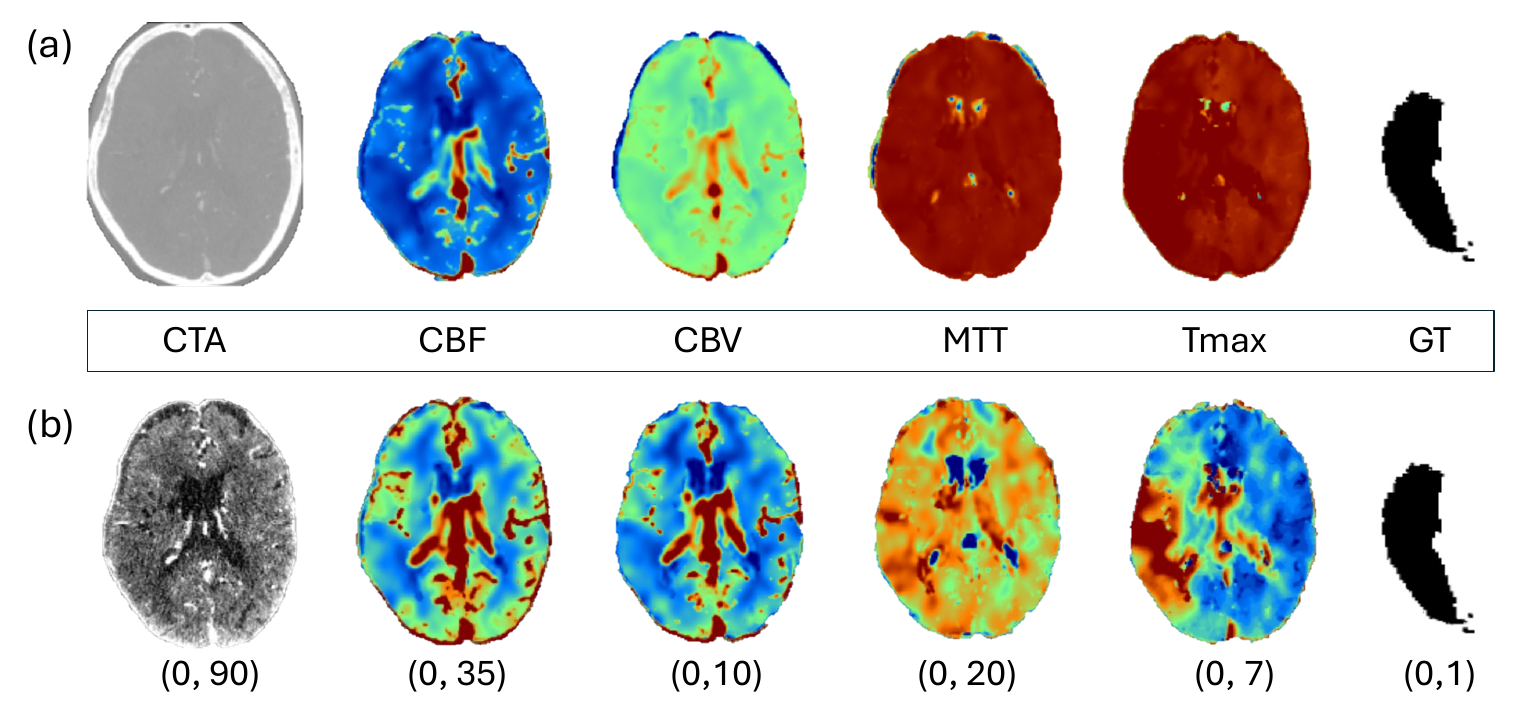}
    \caption{Data (a) before and (b) after preprocessing, with windowing bounds shown below b).}
    \label{fig:pre_vs_post}
\end{figure}

To remove extraneous information from CT images, intensity value windowing was applied.  When clinical windowing guidelines were available in the literature, they informed the initial settings, with empirical adjustments made to optimize model performance. The clinically established thresholds used for reference include: CBF~$<$~17~mL/100g/min~\cite{bandera2006cbfthreshold}; CBV~$>$~2~mL/100g~\cite{wintermark2006perfusion}; MTT~$>$~145\% of the contralateral baseline, where the 0--30~HU range captures the full variability~\cite{nukovic2023neuroimaging,alzahrani2023assessing,czap2021overview}; and Tmax~$>$~6~s~\cite{olivot2009optimal}. For CTA, where no formal guidelines exist, the window was manually adjusted to enhance contrast between healthy and ischemic brain regions, following the methodology in~\cite{pulli2012acute}. For inputs with inconclusive clinical thresholds, such as CTA, MTT, and CBF, further widening of the selected windows was performed to improve the visibility of the ischemic lesion. The windowing bounds used for preprocessing in the final submission can be found in Table \ref{tab:ct_window_values}, and an example of a subject's scans before and after windowing are shown in Figure \ref{fig:pre_vs_post}.


\begin{table}[h]
\centering
\resizebox{\linewidth}{!}{
\begin{tabular}{|c|c|c|c|c|c|}
\hline
\textbf{Modality} & CTA (HU) & CBF (mL/100g/min) & CBV (mL/100g) & MTT (s) & Tmax (s) \\
\hline
\textbf{Window Range} & (0, 90) & (0, 35) & (0, 10) & (0, 20) & (0, 7) \\
\hline
\end{tabular}
}
\caption{Windowing ranges for each CT modality used as input to the nnU-Net segmentation model for the final submission to the ISLES'24 challenge.}
\label{tab:ct_window_values}
\end{table}

\section{Results}


10-fold cross validation was performed, and the fold with the highest validation Dice score was selected for submission. On this fold, the standard pipeline with only Z-score normalization and [1,99] percentile windowing achieved a Dice score of 21.8\%, custom windowing alone improved performance to 31.0\%, and our submission combining custom windowing with Z-score normalization further increased the validation Dice score to \textbf{31.8\%}.

\begin{table}[H]
\centering
\begin{tabular}{|l|c|}
\hline
\textbf{Preprocessing} & \textbf{Dice Score} \\
\hline
Only Z-score Normalization & 0.218 \\
Windowing & 0.310 \\
\textbf{Windowing + Histogram Equalization} & \textbf{0.318} \\
\hline
\end{tabular}
\caption{Effect of different preprocessing strategies on Dice score for the best-performing fold of 10-fold cross-validation.}
\label{tab:preprocessing_dice}
\end{table}

\section{Challenge Submission Results}

\begin{table}[h]
\centering
\begin{adjustbox}{width=0.9\linewidth}
\begin{tabular}{|c|c|c|c|c|}
\hline
Team & Dice (\%) $\uparrow$ & AVD $\downarrow$ & F1 (\%) $\uparrow$ & ALCD $\downarrow$ \\
\hline
\textbf{Kurtlab (Ours)} & \textbf{28.50 (21.27)} & \textbf{21.23 (37.22)} & \textcolor{blue}{14.39 (21.19)} & \textcolor{blue}{7.18 (7.67)} \\
AMC-Axolotls & \textcolor{blue}{26.27 (24.73)} & \textcolor{blue}{21.31 (35.23)} & \textbf{14.94 (25.12)} & 7.66 (7.94) \\
Ninjas & 25.46 (19.08) & 26.29 (39.73) & 9.92 (13.46) & \textbf{5.98 (6.46)} \\
\hline
\end{tabular}
\end{adjustbox}
\caption{Test set evaluation metrics for the top 3 entries in the ISLES'24 leaderboard, reported as mean (standard deviation). Metrics shown are: Dice coefficient (Dice), Absolute volume difference 
 (AVD), F1-score, and Absolute lesion count difference (ALCD). Per column, \textbf{Bold} = best, \textcolor{blue}{blue} = second best. }
\label{table:test_metrics_leaderboard}
\end{table}

\section{Analysis}
Compared to standard Z-score normalization with [1,99] percentile windowing, 10-fold cross-validation results show that custom windowing improves CT-based segmentation by 10 Dice percentage points in the best-performing fold. This is likely because excluded regions in standard preprocessing contain mostly skull, and relevant tissue often falls within a narrow band in the [1,99] range. While our proposed method improves performance, it still exhibits large test metric variance (Tab \ref{table:test_metrics_leaderboard}), suggesting robustness should be improved in future works.

\section{Conclusion}
The presented analysis shows that standard preprocessing pipelines, such as those used in nnU-Net for CT scans, are insufficient to segment ischemic stroke lesions in CT. Here, we share the tailored preprocessing steps from our winning solution to the ISLES'24 challenge, which increase the visibility of tissues of interest and improve performance. Further studies in this direction will involve improving robustness of the segmentation approach by implementing clinical priors beyond windowing in the preprocessing pipeline.

\newpage
\midlacknowledgments{The work of Juampablo Heras Rivera was partially supported by the U.S. Department of Energy Computational Science Graduate Fellowship under Award Number DE-SC0024386.}

\bibliography{midl-samplebibliography}

\end{document}